\documentstyle[12pt,iopl12]{ioplppt}

\author{}

\article{{\bf Classical and Quantum Gravity}}{Analytic description of singularities in Gowdy spacetimes}
\author{Satyanad Kichenassamy}
\address{Max-Planck-Institut f\"ur Mathematik in den Naturwissenschaften,
         Inselstra\ss e 22--26, 04103 Leipzig, Germany.
         E-mail: {\tt kichenas\char'100 mis.mpg.de}.}
\author{Alan D. Rendall}
\address{Max-Planck-Institut f\"ur Gravitationsphysik,
         Schlaatzweg 1, 14473 Potsdam, Germany.
         E-mail: {\tt rendall@aei-potsdam.mpg.de}.}
\date{to appear in {\bf Classical and Quantum Gravity}}
\jl{6}

\abstract
We use the Fuchsian algorithm to construct singular solutions of
Einstein's equations which belong to the class of Gowdy spacetimes. The
solutions have the maximum number of arbitrary functions. Special cases
correspond to polarized, or other known solutions. The method provides
precise asymptotics at the singularity, which is Kasner-like. All of these
solutions are asymptotically velocity-dominated. The results account for
the fact that solutions with velocity parameter uniformly greater than one
are not observed numerically. They also provide a justification of formal
expansions proposed by Grubi\v si\'c and Moncrief.
\endabstract

\newcommand{\pa}{\partial}
\newcommand{\ep}{\varepsilon}
\newcommand{\ph}{\varphi}
\newcommand{\eb}{\mbox{{\bf e}}}

\newtheorem{th}{Theorem}
\renewcommand{\today}{}

\begin{document}

\maketitle

\section{Introduction}

The singularity theorems of Hawking and Penrose (see \cite{h-e}) show
that solutions of Einstein's equations are ``non-continuable'' under
rather general conditions, but do not provide very specific
information on the structure of singularities. This motivated several
attempts to try and provide an analytical description of singularities
of solutions of Einstein's equations. Our approach in this paper is to
try and determine how to perturb known exact solutions and to decide
whether or not the type of singularity they possess is representative
of the behavior of more general solutions.

There is a technique which provides precisely this type of
information for rather general classes of partial differential equations:
the Fuchsian algorithm. It consists in constructing singular solutions
with a large number of arbitrary functions by considering the equation
satisfied by a rescaled unknown, which represents in fact the `regular
part' of the solution. This new unknown satisfies a {\em Fuchsian
PDE}, i.e., a system of the form
\[
t{\pa \vec{u} \over \pa t} + A\vec{u} =
f(t,x_1,\dots,x_n,\vec{u},\vec{u}_x),
\]
where $A$ is a square matrix and $f$ vanishes like some power of $t$
as $t\to 0$.
A general introduction to this algorithm, with several applications can be
found in \cite{nlw,syd}, and a brief presentation is given in section
2 below. We just note here that non-singular solutions can also
be constructed by the Fuchsian algorithm. In fact, the Cauchy problem
itself reduces to a very special case of the method.

We prove in this paper that the Fuchsian algorithm applies to Einstein's
vacuum equations for Gowdy spacetimes, and establishes the existence of a
family of solutions depending on the maximal number of arbitrary
functions, namely four, in the `low-velocity' case, whose definition is
recalled below. When one of these functions is constant, the solution
actually extends to the `high-velocity' case as well. We will refer to the
former solutions as `generic' and to the latter as `non-generic.' Earlier
exact solutions are obtained by specializing the arbitrary functions in
the solutions of this paper.

In both cases, the solutions are `asymptotically velocity-dominated'
(AVD) in the sense of Eardley, Liang and Sachs \cite{e-l-s}, and
precise asymptotics at the singularity are given. The reduction to
Fuchsian form actually provides a mechanism whereby inhomogeneous
solutions can become AVD in the neighborhood of such a singularity.
The results explain the paradoxical features of numerical computations
described next.

\subsection{Earlier results}

{\bf T}$^3\times${\bf R} Gowdy spacetimes \cite{g} have spacelike slices,
homeomorphic to the three-torus, on which a U(1)$\times$U(1)
isometry group acts. It is convenient to take as time coordinate the area
$t$ of the orbits of this two-dimensional group; the spacetime
corresponds to the region $t>0$. The metric then takes the form
\[
ds^2=e^{\lambda/2}t^{-1/2}(-dt^2+dx^2)+t[e^{-Z}(dy+X\,dz)^2
                          +e^{Z}dz^2],
\]
where $\lambda$, $X$ and $Z$ are functions of $t$ and $x$ only, and are
periodic of period $2\pi$ with respect to $x$. We also let
\[
D=t\pa_t.
\]

\smallskip

{\bf Form of the equations.}
With the above conventions, the equations take the form:
\begin{eqnarray}
\label{eq}
D^2 X - t^2 X_{xx} & = & 2(DX\,DZ-t^2 X_x Z_x); \\
D^2 Z - t^2 Z_{xx} & = & -e^{-2Z}((DX)^2 - t^2 X_x^2)\\
\lambda_x              & = & 2(Z_xDZ+e^{-2Z}X_xDX);  \nonumber \\
D\lambda               & = & (DZ)^2+t^2Z_x^2+e^{-2Z}((DX)^2+t^2X_x^2),
                         \nonumber
\end{eqnarray}
where subscripts denote derivatives.

The last two equations arise respectively from the momentum and
Hamiltonian constraints. It suffices to solve the first two equations, and
we therefore focus on them from now on. Of course, one should also ensure
that the integral of $\lambda_x$ from 0 to $2\pi$ vanishes.

If $X=Z=0$, we recover a metric equivalent to the (2/3,2/3,$-1/3$)
Kasner solution. Other Kasner solutions are recovered for $X=0$
and $Z=k \ln t$; the corresponding Kasner exponents are
$(k^2-1)/(k^2+3)$, $2(1-k)/(k^2+3)$ and $2(1+k)/(k^2+3)$.
The equations for $X$ and $Z$ are often interpreted as expressing that
$(X,Z)$ generates a \lq harmonic-like' map from 1+1 Minkowski space with values
in hyperbolic space with the metric
\[
dZ^2+e^{-2Z}dX^2.
\]
The usual Cartesian coordinates on the Poincar\'e model of hyperbolic
space are $X$ and $Y=e^{Z}$, so that the metric coincides with the
familiar expression $(dX^2+dY^2)/Y^2$.

It is occasionally useful to use polar coordinates $(w,\phi)$ on the
hyperbolic space, so that the metric on the target space is
\[
dw^2+\sinh^2w\,d\phi^2.
\]
The equations for $w$ and $\phi$ then take the form
\begin{eqnarray}
\label{eq-2}
D^2w - t^2\pa_{xx}w       & = & {1\over 2}\sinh 2w [(Dw)^2-t^2w_x^2]; \\
D^2\phi - t^2\pa_{xx}\phi & = & -2\coth w [DwD\phi-t^2 w_x\phi_x].
\end{eqnarray}

For fixed $t$, the solution represents a {\em loop} in hyperbolic space.

For extensive references on Gowdy spacetimes, see \cite{c,c-i-m,b-m,g-m2}.

\smallskip

{\bf Exact solutions.}
Both sets of equations can be solved exactly if we seek solutions
independent of $x$. In terms of the $(X,Z)$ variables for example, these
solutions have leading behavior of the form
\[
Z\sim k \ln t + O(1);\qquad X = O(1),
\]
where $k$ is a positive constant,
and represent solutions in which the loop degenerates to a point which
follows a geodesic and tends to a point at infinity in hyperbolic space.

Motivated by this, it was suggested that a general solution corresponds to
a loop, each of whose points asymptotically follows a geodesic and tends
to some point at infinity in hyperbolic space. This regime
is called the `geodesic loop approximation.'

This is borne out in the case of the `circular loop' which corresponds to
$\phi(x)=nx$, $n=1$, 2,\dots, and $w$ independent of $x$.

However, the only case in which this behavior could be established for
solutions containing an arbitrary function of $x$ was the `polarized'
case, defined by the condition $X\equiv 0$ (see \cite{c-i-m} and its
references). The equation for $Z$ is then a linear Euler-Poisson-Darboux
equation, the general solution of which can be represented explicitly in
terms of Bessel functions; this fact does not necessarily make the
investigation of singularities straightforward, see [3].
This provides a family of solutions involving two arbitrary functions.
These solutions have
\[
Z\sim k \ln t + O(1),
\]
where $k$ now depends on $x$ and can be arbitrary.

Numerical computations suggest more complicated behavior in the full
nonlinear system for $X$ and $Z$ \cite{b-m}. Indeed, if one monitors the
`velocity' $v(x,t)=\sqrt{(DZ)^2+\exp(-2Z)(DX)^2}$, which should tend to
$|k(x)|$, one finds that it is not possible to find solutions which
satisfy $v>1$ on any interval as $t\to 0$. Even if one starts out with
$v>1$ and solves towards $t=0$, the parameter $v$ dwindles to values less
than 1, except for some sharp spikes located near places where $X_x=0$,
and which eventually disappear at any fixed resolution. They may persist
longer at higher resolutions. Solutions such that $v<1$ are called
`low-velocity,' and others are called `high-velocity.' A formal asymptotic
computation, proposed in \cite{g-m2}, also suggests that the low-velocity
case allows asymptotics that would not be available in the high-velocity
case. This expansion is obtained by introducing a parameter $\eta$
in front of the spatial derivative terms in the equations, and expanding
the solution in powers of $\eta$.

Note that solutions with $k$ positive and negative are qualitatively quite
different, even though they would have the same value for $v$.

Since the numerical computations we wish to account for were performed in
the $(X,Z)$ variables, we will focus on them. However, we will briefly
mention what happens in the $(w,\phi)$ variables, since the circular loop
is then more simply described.

The problem can be summarized as follows: if the geodesic loop
approximation is valid, $v$ approaches $|k|$. We
therefore need a mechanism which forces $|k|<1$---but if $v$ must be
smaller than 1, how do we account for the polarized solutions? Also,
should we restrict ourselves to $k>0$, given that the numerics do not
give information on the sign of $k$?

\smallskip

{\bf Results.}
Our results account for the various types of behavior observed on
numerical and special solutions by exhibiting a solution with the
maximum number of `degrees of freedom,' and which, under specialization,
reproduces the main features listed above. We describe these results first
for $k>0$.

When $k$ is positive, we first define new unknowns
$u(x,t)$ and $v(x,t)$ by the relations
\begin{eqnarray}
\label{def}
Z(x,t) & = & k(x) \ln t + \ph(x) + t^\ep u(x,t);\\
\label{deff}
X(x,t) & = & X_0(x) + t^{2k(x)}(\psi(x) + v(x,t)),
\end{eqnarray}
where $\ep$ is a small positive constant to be chosen later.
The objective is to construct solutions of the form (\ref{def}-\ref{deff}),
where $u$ and $v$ tend to zero as $t$ tends to zero.
If $0<k<1$, the periodicity condition $\oint \lambda_xdx=0$ is equivalent to
\begin{equation}
\label{p-cond}
\int_0^{2\pi} k(\ph_x+2 X_{0x} \psi e^{-2\ph})dx = 0,
\end{equation}
which we assume from now on. If $k>1$, we  will require in addition
that $X_{0x}\equiv 0$, for reasons described later. In both cases, we
find that $\lambda = k^2 \ln t + O(1)$ as $t\to 0$.

We then prove that, upon substitution of (5)-(6) into (1)-(2), we
obtain a Fuchsian equation for $(u,v)$, in which the right-hand side
may contain positive and negative powers of $t$, as well as
logarithmic terms. If there are only positive powers of $t$, possibly
multiplied by powers of $\ln t$, we prove an existence theorem which
ensures that there are actual solutions of this form in which $u$ and
$v$ tend to zero. In fact, one can derive iteratively a full expansion
of the solution near the singularity at $t=0$. We prove that there are
only positive powers of $t$ in two cases:
\begin{itemize}
\item if $k$ lies strictly between 0 and 1; this provides a `generic'
solution involving four arbitrary functions of $x$, namely $k$, $X_0$,
$\ph$ and $\psi$.
\item if $k>0$ and $X_0$ is independent of $x$; this provides a solution
involving only three functions of $x$ and one constant. This case includes
both the
$x$-independent solutions and the polarized solutions, and explains why
these cases do not lead to a restriction on $k$.
\end{itemize}
The fact that high-velocity is allowed when $X_0$ is constant is to be
compared with the numerical results which show spikes when $X_x=0$.

If $k$ is negative, one can proceed in a similar manner, except that one should
start with
\begin{eqnarray}
\label{def-}
Z & = & k(x)\ln t + \ph(x) + t^\ep u(x,t);\\
X & = & X_0(x) + t^\ep v(x,t),
\end{eqnarray}
where $k$, $\ph$ and $X_0$ are arbitrary functions. In fact, one can
generate solutions with negative $k$ from solutions with positive $k$.
Indeed, if $(X,Z)$ is any solution of the Gowdy equations, so is
$(\tilde{X}, \tilde{Z})$, where
\[
\tilde{X} = {X\over X^2 + Y^2}, \quad
\tilde{Z} = \ln {Y\over X^2 + Y^2},
\]
with $Y=e^Z$ as before. This corresponds to an inversion in the Poincar\'e
half-plane.

\smallskip

Our existence results can actually be applied in two different ways to
the problem. One is to assume the arbitrary functions to be analytic and
$2\pi$-periodic, and to produce solutions which are periodic in $x$. One
can also use the results to produce solutions which are only defined near
some value of $x$. This is useful for cases when the solution is not
conveniently represented in the $(X,Z)$ coordinates, in which one of the
points at infinity in hyperbolic space plays a distinguished role. In such
cases, one can patch local solutions obtained from several local charts in
hyperbolic space.

\subsection{Organization of the paper}

\hskip\parindent%
Section 2 presents a brief introduction to Fuchsian techniques.

Section 3 is devoted to the reduction of the basic equations to Fuchsian
form, and shows how the distinction between low- and high-velocity arises
naturally from the Fuchsian algorithm (theorems 1 and 2).

Section 4 proves the existence result (theorem 3) which
produces the above solutions. It also shows the impact of the rigorous
results on formal asymptotics.


\section{Introduction to Fuchsian techniques}

We briefly review the main features of Fuchsian methods as they are
relevant to our results. The main advantages of these techniques are:
\begin{enumerate}
\item Fuchsian reduction provides an asymptotic representation of
  singular solutions of fairly general partial differential equations.

\item The arbitrary functions in this representation generalize the
  Cauchy data, in the sense that their knowledge is equivalent to the
  knowledge of the full solution. The Cauchy problem is itself a
  special case of the Fuchsian algorithm.

\item The reduction of a PDE to Fuchsian form explains why solutions
  should become AVD, i.e., how the spatial derivative terms can become
  less important than the temporal derivatives near singularities,
  even though the solution is genuinely inhomogeneous.
\end{enumerate}

The starting-point is a reinterpretation of the solution of the Cauchy
problem for, say, a second-order equation
\[
F[u] = 0.
\]
The geometric nature of the unknown is not important for the following
discussion.  Solving the Cauchy problem amounts to showing that the
solution is determined by the first two terms of its Taylor expansion:
\[
u = u^{(0)} +t u^{(1)} + \dots.
\]
One can think of $u^{(0)}$ and $u^{(1)}$ as prescribed on the initial surface
$\{t=0\}$. This statement does not require any information on the
geometric meaning of the unknown $u$, which may be a scalar or a
tensor for instance.

However, this representation may fail if the solution presents
singularities. The Fuchsian approach seeks an alternative
representation near singularities, in a form such as
\[
u = t^\nu(u^{(0)} +t u^{(1)} + \dots).
\]
There are several issues that need to be dealt with if one seeks such
a solution:
\begin{enumerate}
\item How do we construct such a series formally to all orders? The
  question is far from trivial because any amount of inhomogeneity for
  example can force the appearance of logarithmic terms at arbitrarily
  high orders. Furthermore, the arbitrary terms in the series can
  occur at very high orders even if the equation is only of second order.
\item How do we know there is one solution corresponding to this
  expansion, rather than infinitely many solutions differing by
  exponentially small corrections?
\item How restrictive is it to start with power behavior: in
  particular, is logarithmic behavior allowed?
\end{enumerate}
Once this has been done, the formal series can be used much in the
same way as an exact solution would.

It turns out that all of these issues can be addressed simultaneously
by reducing the given equation to a Fuchsian PDE by the following
program:

First, identify the leading terms.  This requires being able to find
an expression $a(x^q)$ in the coordinates $x^q$ such that, upon
substitution of $a$ into the equation, the most singular terms cancel
each other.

Second, define a renormalized regular part $v$ by setting, typically,
\[
u = a + t^m v.
\]
If $a$ is a formal solution up to order $k$, it is
reasonable to set $m=k+\varepsilon$. If the structure of logarithmic
terms is made explicit, one can also specify the dependence of $v$ on
logarithmic variables, as in \cite{nlw,gks}. There is a considerable variety in
the choices of the renormalized part $v$, and the list of possible
cases where these ideas apply seems to be growing.

Third, obtain the equation for $v$. It is important to ensure, by
introducing derivatives of $v$ as additional variables if necessary,
that one is left with a {\em Fuchsian system}, that is, one of the form
\[
t{\pa v\over \pa t} + A v = t^\varepsilon f(t,x,v,\partial_\alpha v),
\]
where $A$ is a matrix, which could depend on spatial variables, but
should be independent of $t$---otherwise we could incorporate the time
dependence into $f$. $\partial_\alpha v$ stands for first order spatial
derivatives; a second-order equation is converted to such a form by
adding derivatives of the unknowns as additional unknowns. In
general, $f$ can be assumed to be analytic in all of its arguments
{\em except} $t$, because $a$ may contain logarithms or other more
complicated expressions.

Fuchsian PDE are a generalization of linear ordinary differential
equations with a regular, or Fuchsian singularity at $t=0$, such as the
Bessel or hypergeometric equations.

Once this reduction has been accomplished, general results on Fuchsian
equations give us the desired results, intuitively because
the equation can be thought of as a perturbation of the case when
$f=0$. The initial-value problem for such equations can be solved in
the non-analytic as well as the analytic case \cite{hs}.

The Fuchsian form has several advantages, in addition to being the
one which allows one to construct and validate the expansions in the
first place:
\begin{enumerate}
\item It makes AVD behavior natural, because the spatial derivative
  terms appear only in $f$, which is preceded by a positive power of
  $t$. We therefore expect spatial derivative terms to be switched off
  at leading order, but to contribute at higher order. By contrast,
  the term $t\partial_t v$ behaves like a term of order zero, because
  it transforms any power $t^j$ into a multiple of itself (namely
  $jt^j$).
\item It is invariant under restricted changes of coordinates which
  preserve the set $t=0$: if we change $(t,x^\alpha)$ into
  $(t',x^{\prime \alpha})$, it suffices to require that $t'/t$ be
  bounded away from zero and independent of $x^\alpha$ near $t=0$. One
  can even allow non-smooth changes of coordinates such as
  $t'=t^\alpha$. Further generalizations are possible.
\item It is invariant under `peel-off': for instance, if we write
  $v=v^{(0)}+tw$, and assume for simplicity that $\varepsilon =1$, we
  find that $w$ solves a Fuchsian system with $A$ replaced by $A+1$. A
  more general property of this kind can be found in \cite{gks}. This
  explains why the Fuchsian form is adapted to the construction of
  formal solutions as well as to their justification.
\item It can be used to generate the formal expansion systematically:
  assume the solution is known to some order $k$. Substitute into $f$,
  and call $g$ the result; now solve the resulting equation
  $t\partial_t v + A v = t^\varepsilon g$ for $v$. It is easy to see
  that the result will contain corrections of order higher than $k$.
  This method is useful if the exact form of the solution is unknown,
  or if it is very complicated.
\end{enumerate}

\smallskip

Let us now turn to examples.

{\bf 1. The Cauchy problem.}
The Cauchy problem can always be thought of as an initial-value
problem for a first-order system
\[
{\pa u\over \pa t} = f(t,x,u,\partial_x u),
\]
where $x=(x^\alpha)$ stands for several space variables, and $f$ is
analytic in all its arguments to fix ideas. For instance, in the case
of Einstein's equations in harmonic coordinates, $u$ represents the
list of all the components of the metric as well as their first time
derivatives.

Let us now take as principal part $a$ the initial condition $u^{(0)}$,
and write
\[
u=u^{(0)} + tv.
\]
If we insert this into $f$, we find that all of the $v$-dependent
terms must contain a positive power of $t$. In other words,
\[
f=f^{(0)}+tg(t,x,v,\partial_x v),
\]
where $f^{(0)}=f(0,x,u^{(0)},\partial_x u^{(0)})$.
The equation for $v$ is therefore
\[
t{\partial v\over \partial t} + v = f^{(0)}+tg,
\]
which is a Fuchsian equation for $(v-f^{(0)})$, with $A=1$. The
existence of solutions of Fuchsian systems ensures in this case that
one can solve the initial-value problem. To recover a solution of
Einstein's equations, one needs to handle the propagation of the
constraints separately, as usual.

\smallskip

{\bf 2. A nonlinear ODE.}
Consider the equation
\[
u_{tt} = u^2,
\]
where subscripts denote derivatives, and $u=u(t)$ is a scalar.

Let us try to find a leading part of the form $u\sim a t^s$ with
$a\neq 0$. The left-hand side is then $\sim as(s-1)t^{s-2}$ and the
right-hand side is $\sim a^2 t^{2s}$. If $s(s-1)=0$, it means that we
are dealing with a Cauchy problem: $u\approx a+u_1 t+\dots$ if
$s=0$, and $u\approx a t+u_2 t^2\dots$ if $s=1$. We therefore assume
$s(s-1)\neq 0$. It is then necessary for the two sides to balance each
other as $t\to 0$, which means that we need
\[
s-2=2s \mbox{ and } s(s-1) = a.
\]
This means that $s=-2$ and $a=6$. The principal part is $6/t^2$, and the
first step is complete.

For the second step, let us define the renormalized unknown $v$ by
\[
u=t^{-2}(6 + v t).
\]

Finally, let us write the equation for $v$. We find
\[
(D-5)(D+2)v = tv^2,
\]
where $D=td/dt$.  This is a Fuchsian equation of second order, which
can be converted into a first-order Fuchsian system by introducing
$(v,Dv)$ as two-component unknown. This would lead to an equation
where the eigenvalues of $A$ are $2$ and $-5$.

The knowledge of the eigenvalue $-5$ combined with general properties
of Fuchsian systems ensures that there is a complete formal solution
for $v$ where the coefficient of $t^5$ in the expansion of $v$ is
arbitrary. One can convince oneself of this fact by direct
substitution, but this is often cumbersome, because of the
need to compute a formal solution to sixth order in this case.
In general, the expansion of $v$ contains also powers of $t\ln t$, but
they are not necessary for this simple example.

The same method applies to any equation of the form
\[
u_{tt} = u^2 + c_1(t)u + c_0(t) + c_{-1}(t)u^{-1} +\dots
\]
and yields a convergent series solution
\[
u(t)=t^{-2}\sum_{j,k} u_{j,k}t^j(t\ln t)^k,
\]
which is entirely determined once the coefficient $u_{6,0}$ is
prescribed. The translates $u(t-t_0)$ of this solution form a
two-parameter family of solutions, parametrized by $(u_{6,0},t_0)$,
which is stable under perturbations (i.e., `generic'). It is possible
to show that the other eigenvalue of $A$, namely 2, is related to the
variation of the parameter $t_0$, although we do not dwell on this
point.

Logarithmic terms are {\em not} due to logarithms in the equation
itself. For instance, the equation $u_{tt}=u^2+t^2$ has no solution
which is free of logarithms.

\smallskip

{\bf 3. The Euler-Poisson-Darboux equation.}
As an example of a linear Fuchsian PDE, let us consider the
Euler-Poisson-Darboux (EPD) equation
\[
u_{tt} - {\lambda -1\over t}u_t =u_{xx}+u_{yy},
\]
in two space variables to fix ideas. This equation has a variety of
uses, from the solution of the wave equation in Minkowski space to
computer vision. In particular, the Einstein equations in the
`polarized' Gowdy spacetime (i.e., when $X=0$) reduce to the above
equation with only one space variable, and with $\lambda=0$.

To reduce it into Fuchsian form, one may let introduce new unknowns: $v=u$,
$v_0=tu_t$, $v_1=tu_x$ and $v_2=tu_y$ (numerical subscripts do
not denote derivatives). One then finds the system
\[
\left\{
\begin{array}{rcl}
t\pa_t v   - v_0          & = &  0\\
t\pa_t v_0 - \lambda v_0  & = &  t\pa_xv_1+t\pa_yv_2\\
t\pa_t v_1                & = &  t\pa_x(v+v_0)\\
t\pa_t v_2                & = &  t\pa_y(v+v_0).
\end{array}
\right.
\]
The general solution can in this case be computed explicitly using the
Fourier transform (or Fourier series in a finite domain) in terms of
Bessel functions. The solution has the form $U+V\ln t$, where $U$ and
$V$ are series in $t$ and do not involve logarithms.

Fuchsian reduction applies directly to non-linear perturbations of the EPD
equation. However, the non-linearity causes the appearance of products
of logarithms. The Fuchsian algorithm, by ensuring that the solutions
are actually functions of $t$ and $t\ln t$, guarantees that the
singularity of the logarithm is always compensated by powers of $t$.
\smallskip

{\em Remark:}
There are cases when it is useful to make a change of
time variable.  Consider an example such as
\[
(t\pa_t)^2u - tu_{xx}=0.
\]
If we let $(v,v_0,v_1)=(u,tu_t,tu_x)$, we obtain the system
\[
\left\{
\begin{array}{rcl}
t\pa_t v   & = & v_0\\
t\pa_t v_0 & = & \pa_xv_1\\
t\pa_t v_1 & = & v_1 + t\pa_xv_0,
\end{array}
\right.
\]
in which the term $\pa_xv_1$ does not have a factor of $t$. We can
nevertheless obviate this problem by letting $t=s^2$. The
original equation then becomes
\[
(s\pa_s)^2u - 4s^2u_{xx}=0;
\]
expanding and dividing through by $s^2$, we recover the
Euler-Poisson-Darboux equation, up to the harmless factor of 4.

\smallskip

{\bf 4. Leading logarithms.}
The first case to be treated by Fuchsian PDE methods actually required
a logarithmic leading term. We merely state the result, as it is
developed extensively elsewhere \cite{hs,inversibilite}. Consider the equation
\[
\eta^{ab}\partial_{ab}u = e^u,
\]
in Minkowski space. This equation admits a Fuchsian reduction with
singularity on any space-like hypersurface $t=\psi(x)$, which is
obtained by applying the above ideas to the equation satisfied by
$e^u$. This generates a family of stable singularities which do not
propagate on characteristic surfaces, since the singularity locus is
space-like. There is a complete expansion of the solution at the
singularity, and it is free of logarithms if and only if the
singularity surface has vanishing scalar curvature (i.e.,
${}^{(3)}R=0$).

\smallskip

To summarize, the Fuchsian approach to singularity formation consists
in three steps: (1) identification of the leading part; (2)
identification of a convenient renormalized unknown; (3) solution of
the Fuchsian system for the new unknown. This technique is now applied
to the Gowdy problem.


\section{Reduction to Fuchsian form}

\subsection{General results}

In this section, we first reduce the Gowdy equations to a second-order
system for $u$ and $v$, which is then converted to a first-order
Fuchsian system. The subscripts $0$, 1 and 2 in this section do {\em
  not} denote derivatives.  The equations now become:
\begin{eqnarray}
(D+\ep)^2u & = & t^{2-\ep}[k_{xx}\ln t + \ph_{xx}+t^\ep u_{xx}]
                                        \nonumber  \\
           &   & \quad\mbox{}-\exp(-2\ph-2t^\ep u)
    \big\{t^{2k-\ep}((D+2k)(v+\psi))^2              \nonumber  \\
           &   & \quad\;\mbox{}-t^{2-2k-\ep}
                   [X_{0x}+t^{2k}(v_x+\psi_x+2k_x(v+\psi)\ln t)]^2
    \big\};
                                              \\
D(D+2k) v  & = & t^{2-2k} X_{0xx} + 2t^\ep(D+\ep)u(D+2k)(v+\psi)
                                        \nonumber  \\
           &   & \;\mbox{}+
      t^2[(v+\psi)_{xx}+4k_x(v_x+\psi_x)\ln t    \nonumber  \\
           &   & \qquad\qquad\mbox{}+
                    (2k_{xx}\ln t + 4k_x^2(\ln t)^2)(v+\psi)]
                                         \nonumber  \\
           &   & \;\mbox{}-
      2t^{2-2k}[X_{0x}+t^{2k}(v_x+\psi_x+k_x(v+\psi)\ln t)]
                                         \nonumber  \\
           &   & \;\mbox{}\times
               [k_x\ln t+\ph_x+t^\ep u_x]. \end{eqnarray}

This second-order system will now be reduced to a first-order system.
To this end, let us introduce the new variables
\[
\vec{u} = (u_0,u_1,u_2,v_0,v_1,v_2) =
(u,Du,tu_x,v,Dv,tv_x).
\]
We then find
\[
\begin{array}{rcl}
Du_0 & = & u_1; \\
Du_1 & = & \mbox{}-2\ep u_1-\ep^2u_0+t^{2-\ep}(k_{xx}\ln t
              +\ph_{xx})+t\pa_x u_2    \\
     &   & \mbox{}-\exp(-2\ph-2t^\ep u_0)
           \big\{  t^{2k-\ep}(v_1+2kv_0+2k\psi)^2-t^{2-2k-\ep}X_{0x}^2  \\
     &   & \quad\mbox{}-2t^{1-\ep}X_{0x}(v_2+t\psi_x+k_x(v_0+\psi) t\ln t)\\
     &   & \quad\mbox{}- t^{2k-\ep}(v_2+t\psi_x+2k_x(v_0+\psi) t\ln t)^2 \big\};
                                                                \\
Du_2 & = & t\pa_x(u_0+u_1);  \\
Dv_0 & = & v_1; \\
Dv_1 & = & -2kv_1 + t^{2-2k}X_{0xx} + t\pa_x (v_2+t\psi_x)
                  + 4k_x(v_2+t\psi_x) t\ln t \\
     &   & \mbox{}+(v_0+\psi)[2k_{xx}t^2\ln t + 4(k_xt\ln t)^2]  \\
     &   & \mbox{}+2t^\ep(v_1+2kv_0+2k\psi)(u_1+\ep u_0)   \\
     &   & \mbox{}-2X_{0x}t^{2-2k}(k_x\ln t+\ph_x+t^\ep\pa_x u_0) \\
     &   & \mbox{}-2t (\pa_x (v_0+\psi)+2k_x(v_0+\psi)\ln t)
                          (k_xt\ln t+t\ph_x+t^\ep u_2);\\
Dv_2 & = & t\pa_x(v_0+v_1);
\end{array}
\]

This system has therefore the form
\begin{equation}
\label{f}
(D+A)\vec{u} = g(t,x,\vec{u},\vec{u}_x),
\end{equation}
where the right-hand side $g$ involves various powers of $t$, possibly
multiplied with logarithms. We will choose $\ep$ so that all of these
terms nevertheless tend to zero as $t$ goes to zero. The low-velocity case
is precisely the one in which it is possible to achieve this without
making any assumptions on the arbitrary functions entering in the system,
namely $k$, $X_0$, $\ph$ and $\psi$.

In fact, the high- and low-velocity cases are now distinguished by the
absence or presence of the terms involving $t^{2-2k}$ (and
$t^{2-2k-\ep}$).  As is clear from the above equations, these terms
disappear precisely if $X_0$ is a constant (i.e., $X_{0x}=0$).

For any positive number $\sigma$, we define the matrix
\[
\sigma^A = \exp(A\ln\sigma):=\sum_{r=0}^\infty {(A\ln\sigma)^r\over r!}.
\]
One checks by inspection that the matrix $A$ has eigenvalues $\ep$, 0,
and $2k$, and that there is a constant $C$ such that
$|\sigma^A|\leq C$ for any $\sigma\in(0,1)$ if $\ep>0$. This can be seen
for instance by reducing $A$ and explicitly computing the matrix
exponential.

Note that this system is of Cauchy-Kowalewska type for $t>0$, and that
the solutions will in fact be analytic in all variables for $t>0$. The
issue is to construct solutions with controlled behaviour as $t\to 0$.

We are interested in solutions of (\ref{f}) which satisfy $\vec{u}=0$ for
$t=0$. Let us check that these solutions have the property that $u_0$ and
$v_0$ solve the original Gowdy system. Since the second and fifth equation
of the system satisfied by $\vec{u}$ are obtained directly from the
second-order system, it suffices to check that $u_1=Du_0$, $v_1=Dv_0$,
$u_2=tu_{0x}$ and $v_2=tv_{0x}$. The first two statements are identical
with the first and fourth equations respectively. As for the last two, we
note that the first and third equations imply
\[
D(u_2-t\pa_x u_0)=t\pa_x(u_0+u_1-Du_0-u_0)=0.
\]
Since $u_2-t\pa_x u_0$ tends to zero as $t\to 0$, it must be identically
zero for all time, as desired. The same argument applies to $v$.

\smallskip

The computations for the case $k<0$ are entirely analogous, and are
therefore omitted.

\smallskip

We now study the low- and high-velocity cases separately.

\subsection{Low-velocity case}

The following theorem gives the existence of a solution depending
on four arbitrary functions in the case when $k$ lies between zero and
one:
\begin{th}
Let $k(x)$, $X_0(x)$, $\phi(x)$ and $\psi(x)$ be real analytic, and
assume $0<k(x)<1$ for $0\leq x\leq 2\pi$.
Then there exists a unique solution of the form (5)--(6), where $u$ and
$v$
tend to zero as $t\to 0$.
\end{th}

{\em Proof:}
By inspection, the vector $\vec{u}$ satisfies a system of the form
(\ref{f}), where $g$ can be written $t^\alpha f$, provided that we
take $\alpha$ and $\ep$ to be small enough. Letting $t=s^m$, we obtain a new
system of the same form, but with $\alpha$ replaced by $m\alpha$. By
taking $\alpha$ large enough, we may therefore assume that we have a
system to which theorem \ref{fex} below applies. The result follows.

\subsection{High-velocity case}

The following theorem gives the existence of a solution depending
on three arbitrary functions in the case when $k$ is only assumed to
be positive, and may take values greater than one. If $k$ is less than
one, we recover the solutions obtained above, but with $X_{0x}=0$:
\begin{th}
Let $k(x)$, $\phi(x)$ and $\psi(x)$ be real analytic, and
assume $X_{0x}=0$ and $k(x)>0$ for $0\leq x\leq 2\pi$.
Then there exists a unique solution of the form (5)--(6), where $u$ and
$v$ tend to zero as $t\to 0$.
\end{th}

{\em Proof:}
Since $X_{0x}$ is now zero, we find that $\vec{u}$ satisfies, if
$\ep>0$, a Fuchsian system of the form (\ref{f}), where $g$ can be
written $t^\alpha f$, provided that we take $\alpha$ and $\ep$ to be
small enough. Letting as before $t=s^m$, we obtain a new system of the
same form, but with $\alpha$ replaced by $m\alpha$. By taking $m$
large enough, we may therefore assume that we have a system to which
theorem \ref{fex} applies. The result follows.

\section{Existence of solutions of Fuchsian systems}

Consider quite generally a Fuchsian system, for a `vector' unknown
$u(x,t)$, of the form
\begin{equation}
  \label{fe}
  (D+A)u = F[u] := tf(t,x,u,u_x).
\end{equation}
We are now dropping the arrow on $u$ for convenience.  In this
equation, $A$ is an analytic matrix near $x=0$, such that
$\|\sigma^A\|\leq C$ for $0<\sigma<1$. The number of space dimensions
is $n$ ($n=1$ for the application to the Gowdy problem). It suffices
that the nonlinearity $f$ preserve analyticity in space and continuity
in time, and depend in a locally Lipschitz manner on $u$ and $u_x$,
i.e., that its partial derivatives with respect to these arguments be
bounded when these arguments are.  To fix ideas, one may assume that
$f$ is a sum of products of analytic functions of $x$, $u$ and $u_x$ by
powers of $t$, $t^{k(x)}$ and $\ln t$. In fact, all one needs is to
ensure the estimate in step 2 below.  In this section, the number of
space variables is arbitrary. We are only interested in positive
values of $t$.
\begin{th}
\label{fex}
The system (\ref{fe}) has exactly one solution which is defined near
$x=0$ and $t=0$, and which is analytic in $x$, continuous in $t$, and
tends to zero as $t\downarrow 0$.
\end{th}

{\em Remark 1:}
The solutions are constructed as the uniform limit of a sequence of
continuous functions which are analytic in $x$. They are classical
solutions as well, by construction. However, by the Cauchy-Kowalewska
theorem, they are also analytic in $t$ away from $t=0$.

\smallskip

{\em Remark 2:}
The solution provided by the theorem will be defined for $x$ in a complex
neighborhood of a subset $\Omega$ of {\bf R}$^n$. This can be applied to
the Gowdy
problem in two different ways: one can take $\Omega$ to be an interval of
length greater than $2\pi$, and note that the solution will be
$2\pi$-periodic if the right-hand side is, since it is given as a
limit of a sequence all of whose terms are periodic. It is this solution
which shows that the `geodesic loop approximation' corresponds to a
generic family of exact solutions in the low-velocity case, and a
non-generic family otherwise. However, one could also take $\Omega$ to be
a small interval of length less than $2\pi$, and generate solutions which
are defined only locally. This second application can itself be useful in
two contexts:
(a) for generalisations of Gowdy spacetimes where the space
variable is unbounded, or compactified in a different fashion; (b) for the
description of `circular loop' type solutions, which correspond to a
solution which depends linearly in the angular coordinate in terms of
polar coordinates on the Poincar\'e half-plane.

\smallskip

{\em Proof:}
Let us begin by defining an operator $H$ which corresponds to the
inversion of $(D+A)$. The proof will consist in showing that the
operator $v\mapsto G[v]:=F[H[v]]$ is a contraction for a suitable norm. Its
fixed point generates a solution $u=H[v]$ to our problem.

Before we jump into the details, let us first motivate the strategy by
examining some of the possible difficulties. For more details on the
history of existence theorems in the complex domain, see Chapter 2 in
\cite{nlw}.

The basic difficulty in achieving a successful iteration is that it is
not clear at all how to build a measure of the size of $u$ (that is, a
function space norm) which remains finite after even one step of the
iteration. The problem is that in order to control $H[v]$, we need to
estimate the spatial derivative of $v$ in terms of a norm which only
involves $v$.  This cannot be remedied by adding information on the
derivatives of $v$ in the definition of the norm: we would then need
to estimate both $H[v]$ and its derivative, in order to have a
well-defined iteration.  In fact, this is an essential problem because
the result would be false if the right-hand side involved second as
well as first derivatives of $u$. Majorant methods are not appropriate
because the nonlinearity $f$ does not have an expansion in powers of
$t$---only in powers of $x$ for fixed $t$.  It is not possible to
estimate the derivative of an analytic function by its values on the
same domain: think of the function $\sqrt{1-z}$ on the unit disk,
which is bounded on $(-1,1)$ even though its derivative is not.
However, by going into the complex domain, it is possible to estimate
the derivatives of an analytic function in terms of its values on the
boundary of a {\em slightly larger domain}.\footnote{In the
  non-analytic case, this problem is gotten around thanks to the
  additional assumption of hyperbolicity, by showing that there are
  expressions which can be estimated {\em as though the right-hand
    side did not involve derivatives of $v$ at all}, see \cite{hs}, ana
  well as Chapter 2 of \cite{nlw} for a broader introduction.} This is
given by Cauchy's theorem, which expresses the value of an analytic
function $\phi$ at any point as a weighted average of its values on
any curve $\gamma$ circling that point once:
\[
\phi(z) ={1\over 2\pi i}\oint_\gamma {\phi(\zeta)d\zeta\over \zeta-z}.
\]
Differentiating with respect to $z$ and taking absolute values, we see
that we have a means of estimating the derivatives of $\phi$ from its
values on a larger domain. However, we must move into the complex
domain to achieve this. The transition to several variables offers no
difficulty, because an analytic function of several variables is
separately analytic in each of its arguments, and it therefore
suffices to apply the above to each variable separately to obtain some
estimate of derivatives---which is all we need. For instance, the
relevant Cauchy integral formula in two variables is simply
\[
\phi(z_1,z_2) ={1\over (2\pi i)^2}\oint\oint
{\phi(\zeta_1,\zeta_2)d\zeta_1 d\zeta_2\over (\zeta_1-z_1)(\zeta_2-z_2)},
\]
where the integration extends over a product of circles:
\[
|\zeta_1-z_1|=r_1 \mbox{ and } |\zeta_2-z_2|=r_2.
\]
The proof below differs from the existence result used in
\cite{gks,nlw} by the fact that the nonlinearity is not analytic
anymore with respect to time. It is therefore necessary to check
carefully that the estimates on $f$ can still be carried out.

\smallskip

We now present the proof of the result.

\smallskip

Throughout the proof, the meaning of the letter $C$ will change from line to
line: it denotes various constants, the specific value of which is not
needed.

We let
\[
H[v] = \int_0^1 \sigma^{A(x)-1}v(\sigma t)\,d\sigma.
\]
It is easily checked that this provides the solution of
\[
(D+A)u=v,
\]
with $u(0)=0$, provided that $v=O(t)$ near $t=0$.

We are ultimately interested in real values of $x$ in some open set
$\Omega$, so that we work in a small complex neighborhood of the real
set $\Omega$. The proof in fact does not depend on the nature or size
of this set.
We also define two norms which will be useful. The $s$-norm of a
function of $x$ is
\[
\|u\|_s = \sup \{ |u(x)| : x\in \mbox{{\bf C}$^n$ and } d(x,\Omega)<s \}.
\]
The $a$-norm of a function of $x$ and $t$ is defined by
\[
|u|_a = \sup
\big\{  {s_0-s\over t} \|u(t)\|_s \sqrt{1-{t\over a(s_0-s)}}
 : t<a(s_0-s) \big\}
\]
Note that this norm allows functions to become unbounded when
$t=a(s_0-s)$. This can be thought of basically as the boundary of the
domain of dependence of the solution. For the reasons indicated
earlier, the iteration would not be well-defined if we had worked
simply with the supremum of the $s$-norm over some time interval.

The objective is to prove that the iteration $u_0=0$, $u_{n+1}=G[u_n]$
is well-defined and converges to a fixed point of $G$, which gives us
the desired solution. This will be acheived by exhibiting a set of
functions which contains zero and on which $G$ is a contraction in the
$a$-norm. Since a contraction has a unique fixed point, we also obtain
uniqueness.

We choose $R>0$ and $s_0$ such that
$\|F[0](t)\|_{s_0}\leq Rt$. This can always be achieved since we are
allowed to take $R$ very large. \smallskip

{\bf Step 1: Estimating $H$.}
Using the definition of $|u|_a$, we find, with the notation
$\rho=\sigma t/a(s_0-s)$,
\begin{displaymath}
  \begin{array}{rcl}
\|H[u](t)\|_s & \leq &
  \displaystyle{   {|u|_a\over s_0-s} \int_0^1 |\sigma^A|{\sigma t\over
\sigma}
         \big(1-{\sigma t\over a(s_0-s)}\big)^{-1/2}\,d\sigma } \\
              & =    &
  \displaystyle{
              {C|u|_a\over s_0-s} \int_0^{t/a(s_0-s)}
                {a(s_0-s)\,d\rho \over \sqrt{1-\rho}}     }      \\
              & \leq  & C_0 a |u|_a.
  \end{array}
\end{displaymath}

\smallskip

{\bf Step 2: Estimating $F$.}
Using Cauchy's integral representation, and the fact that $f$ contains a
factor of $t$, we claim that there is a constant $C_1$ such that
\[
\| F[p]-F[q] \|_{s}(t) \leq {C_1t\over s'-s} \|p-q\|_{s'}
\]
if $s'>s$ and $\|p\|_s$ and $\|q\|_s$ are both less than
$R$; this constraint will be ensured in Step 3 thanks to the argument of
the previous step.

Indeed, $F[p]$ is the product of $t$ by a linear expression in
the gradient of $p$, with coefficients which are Lipschitz functions
of $p$; it is in fact in
the Gowdy case an analytic function of these variables, $x$, and
positive powers of $t$ multiplied by logarithms. The bound on the
$s$-norm ensures that all the partial derivatives of $F$ with respect
to $p$ and $\nabla_x p$ are bounded by some constant $C$.  Therefore,
we have
\[
| F[p]-F[q] | \leq Ct(|p-q|+|\nabla_x p-\nabla_x q|).
\]
We want to estimate the supremum of this expression as $x$ varies so
as to satisfy $\mbox{dist}(x,\Omega)<s$. The first term is clearly
less than or equal to $\|p-q\|_{s}$, and {\em a fortiori} no bigger
than $\|p-q\|_{s'}$. The second is estimated by Cauchy's inequality on
each component. Thus, for the first component, we write
\[
p(x,t)-q(x,t)={1\over 2\pi i}\int_{|z-x_1|=s'-s}
{(p(z,x_2,\dots,t)-q(z,x_2,\dots,t))dz\over z-x_1}.
\]
Differentiating with respect to $x_1$, we find
\begin{eqnarray*}
  |\pa_1(p-q)| & = &
\left| {1\over 2\pi i}\int_{|z-x_1|=s'-s}
{(p(z,x_2,\dots,t)-q(z,x_2,\dots,t))dz\over (z-x_1)^2}\right| \\
& \leq & {1\over 2\pi}\int_{|z-x_1|=s'-s}
{|p(z,x_2,\dots,t)-q(z,x_2,\dots,t)||dz|\over (s'-s)^2}  \\
& \leq & {1\over 2\pi}\|p-q\|_{s'}{2\pi(s'-s)\over (s'-s)^2},
\end{eqnarray*}
which provides the desired estimate for the second term as well.

\smallskip

{\bf Step 3: $G$ is contractive.}
Let us assume in this section that $|u|_a$ and $|v|_a$ are both less
than $R/2C_0a$. We prove that
\[
|G[u]-G[v]|_a\leq C_2 a|u-v|_a.
\]
One should think of $u$ and $v$ as two successive terms $u_n$ and
$u_{n+1}$ in the iterative procedure.  To obtain this inequality, we
first write
\[
G[u]-G[v]=\sum_{j=1}^n F[w_j]-F[w_{j-1}],
\]
where
\[
w_j = \int_0^{j/n} \sigma^{A-1}u(\sigma t)\,d\sigma
     +\int_{j/n}^1 \sigma^{A-1}v(\sigma t)\,d\sigma.
\]
By the arguments of Step 1, we have $\|w_j\|_s< R/2$ for $t<a(s_0-s)$.

We therefore have, using Step 2 with $p=w_j$ and $q=w_{j-1}$,
\[
\|G[u]-G[v]\|_s(t)  \leq
       \sum_{j=1}^n {Ct\over s_j-s} \|w_j-w_{j-1}\|_{s_j}.
\]

Let us choose a sequence of numbers, $s_j=s(j/n)$, where
\[
s(\sigma) = {1\over 2} (s+s_0-{\sigma t\over a}).
\]
We now find
\[
\begin{array}{rcl}
\|w_j-w_{j-1}\|_{s_j} & = &
\big\|
\int_{(j-1)/n}^{j/n} \sigma^{A-1}[u(\sigma t)-v(\sigma t)]\,d\sigma
\big\|_{s_j}         \\
 & \leq & \int_{(j-1)/n}^{j/n} C \|u-v\|_{s(\sigma)}(\sigma
t)\,d\sigma/\sigma \\
 & \leq & \displaystyle{\int_{(j-1)/n}^{j/n} {Ct\over s_0-s(\sigma)}
          { |u-v|_a d\sigma \over \sqrt{1-\sigma t/a(s_0-s(\sigma))} }.}
\end{array}
\]
Letting $n$ tend to infinity, we find the estimate
\[
\|G[u]-G[v]\|_s(t)  \leq
       \int_0^1 C{t^2 |u-v|_a \over (s(\sigma)-s)(s_0-s(\sigma))}
             {d\sigma \over \sqrt{1-\sigma t/a(s_0-s(\sigma)) } }.
\]
We now make the change of variables $\rho=\sigma t/a(s_0-s)$. Note
that
\[
(s(\sigma)-s)(s_0-s(\sigma))={(s_0-s)^2\over 4}(1-\rho^2);\quad
1-{\sigma t\over a(s_0-s(\sigma))} = {1-\rho\over 1+\rho}.
\]
We therefore find
\[
\begin{array}{rcl}
\|G[u]-G[v]\|_s(t) & \leq &
{C a t|u-v|_a\over s_0-s}
       \int_0^{t/a(s_0-s)} {d\rho\over (1-\rho)^{3/2}}  \\
      & \leq &
{C a t|u-v|_a\over s_0-s}\big(1-{t\over a(s_0-s)} \big)^{-1/2} \\
\end{array}
\]
Using the definition of the $a$-norm, we see that we have obtained the
desired estimate.

\smallskip

{\bf Step 4: End of proof.}
Let $u_0=0$ and define inductively $u_n$ by $u_{n+1}=G[u_n]$.
We show that this sequence converges in the $a$-norm if $a$ is
small. Since $\|u_1\|_{s_0}\leq Rt$, we have $|u_1|_a<R/4C_0a$ if $a$ is
small. We may assume $C_2 a<1/2$. It follows by induction that
$|u_{n+1}-u_n|_a\leq 2^{-n}|u_1|_a$,
and $|u_{n+1}|_a < R/2C_0a$, which implies in particular
$\|Hu_n\|_s < R/2$. Therefore all the iterates are well-defined and lie in
the domain in which $G$ is contractive. As a result, the iteration
converges, as desired.

\smallskip

{\bf Impact on formal expansions.}

The expansion of \cite{g-m2} amounts to seeking $X$ and $Z$ as functions of
$(t,\ep x)$, expanding in $\ep$, and then letting $\ep=1$. Its convergence
can therefore be derived from the analyticity of the solutions in $x$.
Note that the reference solution in that paper is slightly more restrictive
than those considered here: they are geodesic loops travelling \lq to the
right' in the Poincar\'e half-plane.

The Fuchsian algorithm provides a different way of generating formal
solutions: by following the existence proof itself. Thus, starting with
$u=0$, we can compute $F[0]$, then solve $(D+A)u_1=F[0]$, which is
a linear ODE in $t$, compute $F[u_1]$, etc. The higher-order
corrections are automatically generated even if their order is not
known in advance.

\smallskip

{\bf Remarks on the nature of the singularity.}

One could check that AVD Gowdy spacetimes with $0<k<1$ or $k>1$ do have a
curvature singularity at $t=0$ by directly computing the Kretschmann
scalar $B:= R_{ijkl}R^{ijkl}$ (for large classes of such spacetimes, see
[2], who uses symbolic manipulation; see also a brief remark in this
direction at the end of [6]). We give a simpler argument which reduces the
issue to the corresponding problem for Kasner spacetimes, where the
answer is classical.

Indeed, consider the orthonormal coframe
\[
(e^{\lambda/4}t^{-1/4}dt,
 e^{\lambda/4}t^{-1/4}dx,
 t^{1/2}e^{-Z/2}(dy+Xdz),
 t^{1/2}e^{Z/2}dz),
\]
and the dual frame $\{\eb_a=\eb_{}^k \partial_k\}$.
One finds, by direct computation, that the Ricci rotation coefficients of
this frame all have the form:
\[
\gamma^a{}_{bc} = C^a{}_{bc} t^{-3(k^2+1)/4} (1+o(1)),
\]
where the leading-order coefficients $C^a{}_{bc}$ are $t$-independent
quantities which involve only $k$:  its derivatives, or
the functions $X_0$, $\varphi$ and $\psi$ do not affect the value of
these coefficients. A similar property holds for the coefficients
$b^c{}_{ab}$ defined by $[\eb_a,\eb_b] = b^c{}_{ab}\eb_c$.
It follows that the product terms in the expression of the frame
components of the curvature tensor are at most $O(t^{-3(k^2+1)/2})$.
As for the Pfaffian derivative terms, it turns out that they are not more
singular, because they are coordinate derivatives multiplied by
appropriate frame components. There are still no $x$-derivatives at
leading order. It follows that the most singular term in $B$ as $t\to 0$
is in fact the same as the one corresponding to $X_0=\varphi=\psi=0$,
and $k=$const., which is the Kasner case.

In extrinsic terms, we may express the result as follows: if $h$ is the
mean curvature of the slices $t=$const., then $B/h^4$ tends to a non-zero
constant for $0<k<1$ or $k>1$, which has the same expression as in the
Kasner case. In particular, $B$ blows up like
$t^{-3(k^2+1)}$, so that we have a curvature singularity, QED.

\smallskip

{\em Remark 1:} It is easy to check that this singularity is reached
in finite proper time by observers with $x=$const., so that this space
is indeed (past) geodesically incomplete.

\smallskip

{\em Remark 2:} There is no change in the leading power of $B$ as $k$ goes
through 1: only the coefficient of the leading term in $B$ vanishes.

\ack
We thank V. Moncrief for numerous helpful discussions during the early
stages of this work. S. K. would like to thank the Max-Planck-Institut in
Potsdam for its warm hospitality during two weeks.

\end{document}